\documentclass{moriond}

\def\be{\begin{equation}}
\def\ee{\end{equation}}
\def\bea{\begin{eqnarray}}
\def\eea{\end{eqnarray}}

\usepackage{siunitx}
\usepackage{graphicx}
\usepackage{caption}
\usepackage{subcaption}
\newcommand{\mps}[1]{\SI{#1}{\meter\per\second}}
\newcommand{\Photo}{}

\begin{document}
\vspace*{4cm}
\title{Fine structure constant measurements in quasar absorption systems}

\author{ D. MILAKOVI{\'C} }

\address{Institute for Fundamental Physics of the Universe, via Beirut 2, 34151 Trieste, Italy}

\maketitle\abstracts{Detecting any evolution of dimensionless in the ratios of physical quantities, such as the fine structure constant, would prove that the Weak Equivalence Principle is violated and lead to a paradigm shift in physics. High resolution spectroscopy of quasar absorption systems can be used to test cosmological variations in time and/or in space. A sample of 300 measurements using data from 8m class optical telescopes provides hints that such variations are indeed present in a form of a spatial dipole across the sky, although systematic effects could dominate. Two recent developments, one in instrumentation and the other in analysis methods, promise to produce a new sample of measurements free from all known systematic effects to test the tentative dipole. }

\section{Introduction}

The fine structure constant, $\alpha = \frac{1}{4\pi\varepsilon_0}\frac{e^2}{\hbar c}$, parametrises the strength of all electromagnetic interactions and, as such, it is considered one of approximately 30 fundamental constants that set the physical properties of our universe. Dirac \cite{Dirac1937} was the first to suggest temporal evolution of fundamental constants, albeit for reasons different from the modern ones \cite{Barrow1998,Barrow2013}. The values of fundamental constants are not predicted by any currently known theory, although there are some recent attempts to do so \cite{Singh2021}. Detection of temporal and/or spatial variations of fundamental constants would immediately disprove Einstein's Equivalence principle and pave the way to a more fundamental theory than the Standard Model of Particle Physics and the $\Lambda$CDM cosmological model. The list of fundamental constants will then have to be revised, consistent with historical trends.
More details on the status of the field of varying constants, including theories which predict such variations, can be found in reviews \cite{Uzan2011,Martins2017}. Experiments aimed at detecting fundamental constant variations are planned for the next generation of large astronomical observatories such as the Extremely Large Telescope \cite{Tamai2022,Marconi2022}.

\begin{figure}
    \centering
    \begin{subfigure}[b]{0.45\textwidth}
    \centering
        \includegraphics[width=\textwidth]{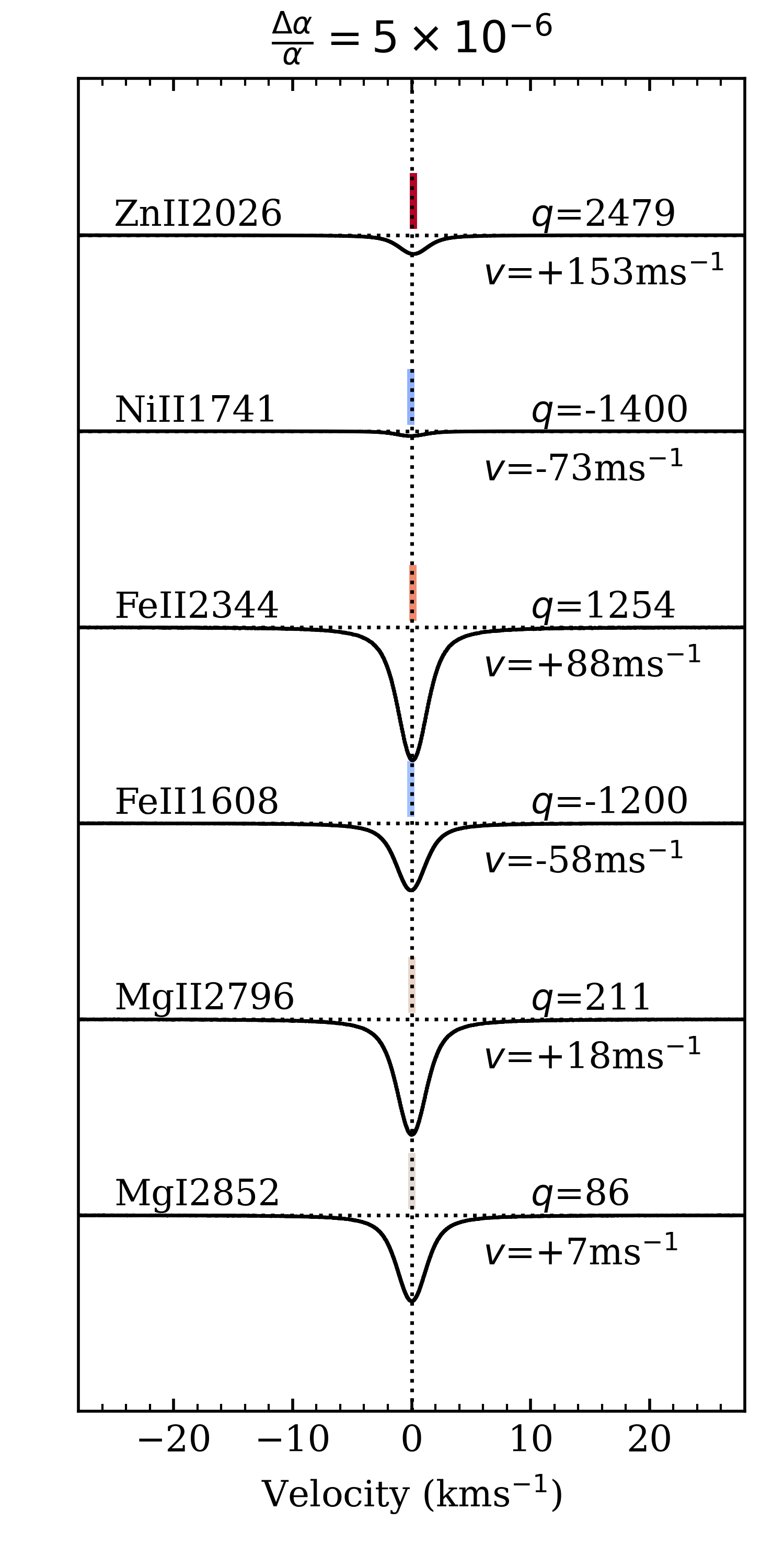}
    \end{subfigure}
    \begin{subfigure}[b]{0.45\textwidth}
    \centering
        \includegraphics[width=\textwidth]{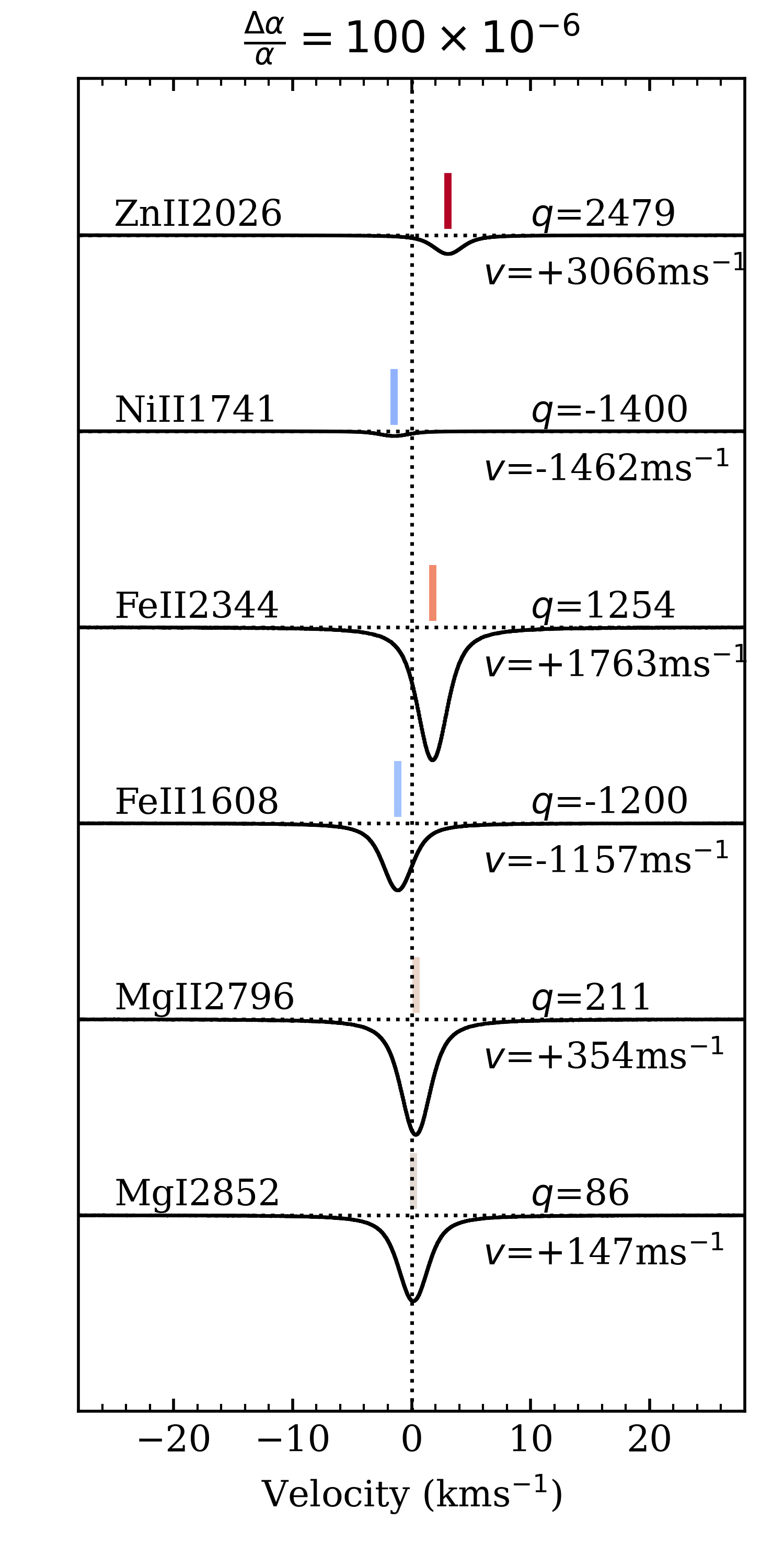}
    \end{subfigure}
    \caption{The observed shifts, $\Delta v$, as per Eq.~\ref{eq:delta_v} for a set of 6 heavy element transitions commonly observed in quasar absorption systems for two values of $\Delta\alpha/\alpha$: $5\times10^{-6}$ (corresponding to the typical uncertainty value of measurements in the literature) and an extreme value of $1\times10^{-4}$ (for illustration of the effect). 
    Information for each plotted transition, including atomic species, rest-frame wavelength (in \AA) and $q$ values (in \SI{}{\per\centi\meter}) are given as insets. The tick marks above each transition indicate the position of the absorption profile centroid and the colour indicates the relative amplitude and the direction of the shift (with respect to zero, i.e.\ $\Delta\alpha/\alpha=0$, indicated by the dotted vertical line). }
    \label{fig:daoa_illustration}
\end{figure}

Any variation in the value of $\alpha$ would be seen as relativistic corrections to the energy levels of electrons around atomic nuclei. The resulting changes in the energy of a given transition can be expressed as a velocity shift with respect to its measured laboratory energy:
\begin{equation}\label{eq:delta_v}
    \frac{ \Delta v } {c} \approx - \frac{q}{2 \omega}\frac{\Delta\alpha}{\alpha},
\end{equation}
where $\Delta v$ is the shift, $c$ is the speed of light, $\omega$ is the transition wave-number and $\Delta\alpha/\alpha = (\alpha_{lab} - \alpha_{obs})/\alpha_{lab}$ with subscripts $obs$ and $lab$ referring to the observed and laboratory values. $q$ quantifies the sensitivity of the transition to $\alpha$ variation and is generally larger for heavy nuclei and for transitions close to the ground state. Measurements in systems exhibiting a number of transitions, spanning as large as possible $\Delta q$ range, forms the basis of the Many Multiplet method \cite{Dzuba1999,Webb1999}. An illustration of velocity shifts for two values, $\Delta\alpha/\alpha=5\times 10^{-6}$ and an extreme value of $\Delta\alpha/\alpha=10^{-4}$, are shown in Figure~\ref{fig:daoa_illustration}.

%For illustration, considering the transitions Fe{\sc ii} $\lambda2383$ ($q=1505$) and Mg{\sc ii} $\lambda2796$ ($q=212$) only, a change of $\Delta\alpha/\alpha=10^{-6}$ would produce a relative shift of $\Delta lambda/\lambda = 6\times10^{-8}$ or \mps{18}. 

Quasars provide means for measurements of $\alpha$ variation at cosmological scales. The Many Multiplet method was applied to a sample of over 300 astronomical systems observed in absorption towards quasars, providing evidence for variation of $\Delta\alpha/\alpha$ with an amplitude of $1\times10^{-5}$ in the form of a dipole across the sky \cite{King2012,Wilczynska2020}. Considering the statistical errors on $\Delta\alpha/\alpha$ only, this result exceeds $4 \sigma$ significance. Of all systematic effects considered, wavelength calibration issues in astronomical spectrographs used for the observations were found to dominate \cite{Griest2010,Rahmani2013,Evans2014}. The impact of calibration issues is a systematic uncertainty comparable to the statistical one \cite{Whitmore2015}, although much smaller than it was initially thought \cite{Dumont2017}.

\section{Recent advancements}
Two recent advancements provide the means to explore the tentative dipole. The first refers to the deployment of novel astronomical instrumentation and the second refers to the development of advanced analysis methods. 

Wavelength calibration methods based on astronomical Laser Frequency Combs \cite{Haensch2006,Steinmetz2008} (LFC) were applied to instruments built for extremely precise spectroscopic measurements such as the HARPS \cite{Mayor2003} and ESPRESSO \cite{Pepe2021} spectrographs installed on the European Southern Observatory's 3.6m telescope and the Very Large Telescope (respectively). The LFC provides \mps{3} accuracy (root-mean square of wavelength calibration residuals) on HARPS \cite{Milakovic2020}, effectively removing the dominant source of instrumental systematics present in all previous measurements. For reasons yet unknown, the LFC on ESPRESSO has only \mps{6} accuracy \cite{Schmidt2021}. Still, the systematic uncertainty from a slightly worse wavelength calibration accuracy is smaller than the typical statistical uncertainty \cite{Schmidt2021,Murphy2022}. More importantly, the LFC can be used to reconstruct the impulse response function of the instrument, the so-called instrumental profile (IP), as discussed later. 

Development of spectral analysis tools, based on genetic (\texttt{GVPFIT} \cite{Bainbridge2017}) and artificial intelligence algorithms (\texttt{AI-VPFIT} \cite{Lee2021aivpfit}) helped establish optimal methods for future measurements. \texttt{AI-VPFIT} combines Monte Carlo methods, genetic algorithms, and information criteria to produce robust and reproducible models of the spectral absorption system in a fully automatic way, without any human decision making and thus bias. Model parameter values, including $\Delta\alpha/\alpha$, are determined through non-linear least-squares optimisation. The Monte Carlo aspect of \texttt{AI-VPFIT} was used to identify previously unknown systematic effects associated with assumptions used in the model-building process \cite{Lee2021nonunique,Webb2022}. Investigation on both simulated and observed data shows that assumptions made on, for example, the gas broadening mechanism or $\Delta\alpha/\alpha$ made during the model-building procedure produces false minima in the highly dimensional $\chi^2$-parameter space. %These are avoided when a more physical model of gas broadening is used, especially when using an information criterion specifically designed to avoid under and over-fitting spectral data \cite{Webb2021}.  
%In essence, modell produce may produce statistically inconsistent $\Delta\alpha/\alpha$ measurements while providing good fits to the data \cite{Lee2021nonunique}.
These problems can be avoided by not imposing any particular assumptions and leaving the data to choose the preferred parameter values \cite{Lee2021nonunique,Webb2021,Webb2022}. All future analyses should be made using \texttt{AI-VPFIT} in this way. 

\section{Quantifying the impact of instrumental profile variations on $\alpha$ measurements}

%\subsection{Instrumental profile reconstruction}\label{sec:ip}
The spectrograph IP is used to convolve the spectrum model before it is compared to the data (in the statistical sense, using  $\chi^2$) when performing parameter optimisation. Providing an incorrect IP shape can result in an incorrect final absorption system model and incorrect $\Delta\alpha/\alpha$. All previous analyses used a Gaussian IP, assuming that the difference between the true and the empirical IP is negliglible. The deployment of astronomical LFCs allows this assumption to be investigated.  

The LFC produces thousands of emission modes with known wavelengths. The wavelengths of the modes are known with relative accuracy of $\Delta \lambda/\lambda = 10^{-12}$ and provide an extremely accurate wavelength calibration for the spectrograph \cite{Steinmetz2008,Wilken2010,Milakovic2020,Probst2020}. The intrinsic width of the LFC modes is approximately 10000 times smaller than the width of the instrument resolution element\footnote{Tilo Steinmetz, private communication} of instruments such as HARPS and ESPRESSO, making them excellent approximations of an monochromatic impulse input. As such, each LFC line image on the detector is a direct representation of the impulse response function, i.e.\ the IP itself. 

\begin{figure}
    \centering
    \includegraphics[width=0.5\columnwidth]{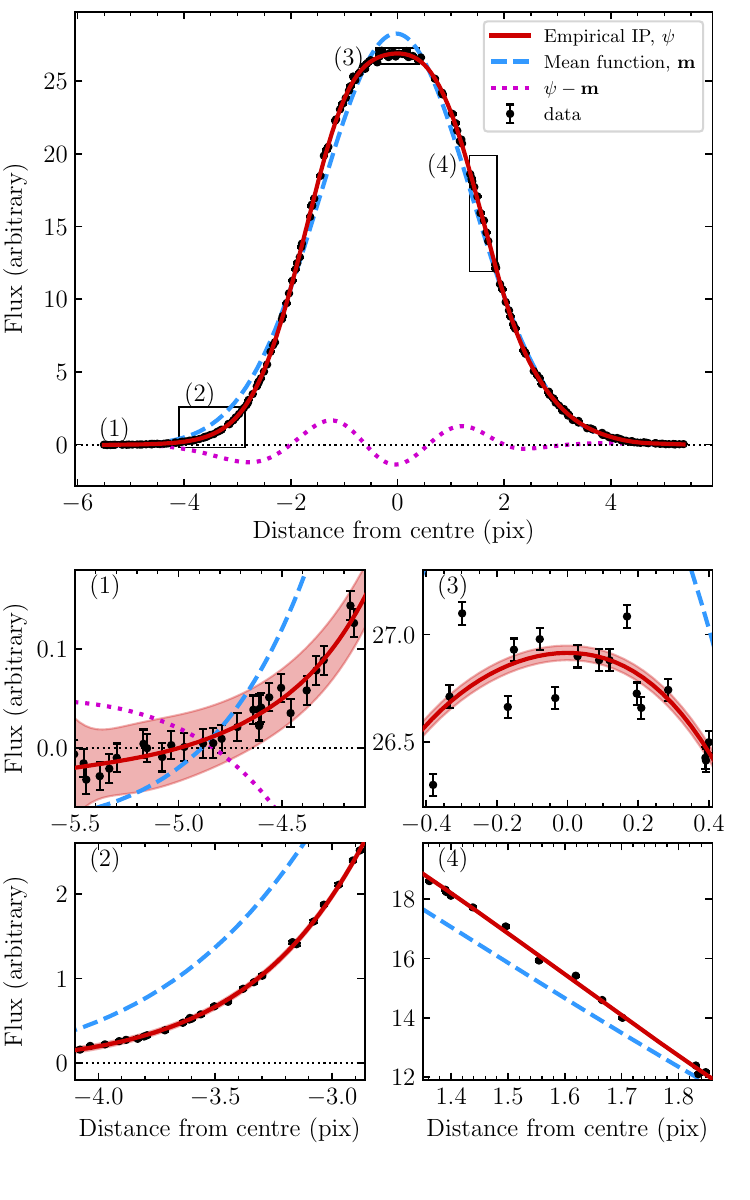}
    \caption{\emph{Top panel}: The black points are the HARPS IP samples with corresponding error bars reconstructed from LFC spectra. The thick red line is our reconstruction of the IP using Gaussian Process regression. Shaded regions show the $1\sigma$ ranges of the model. The dashed blue line is the mean function of the Gaussian process, i.e.\ a Gaussian profile. The purple dotted line is the difference between the Gaussian Process model and the Gaussian mean function. Rectangles indicate regions shown in the bottom panels. \emph{Panels indicated with numbers (1)-(4)}: Zoom-ins of the main panel. The number in the top left corner identifies one rectangle in the main panel. The shaded red bands around the GP model show the $1\sigma$ uncertainty bands. }
    \label{fig:IP_example}
\end{figure}

We have used the LFC lines to reconstruct the IP of HARPS and model it in in a non-parametric way using Gaussian progress regression (Milakovi{\'c} et al., in prep.). Figure \ref{fig:IP_example} shows the an example IP model together with the best fitting Gaussian line profile. Obviously, the empirical IP differs significantly from a simple Gaussian and our non-parametric model presents a significant improvement. We will explore effects associated to IP choice by applying these new IP models to existing HARPS, LFC calibrated, observations of the quasar HE0515$-$4414 for which we already have \texttt{AI-VPFIT} models of the absorption system at $z=1.15$ \cite{Milakovic2021}. These results will be reported in an upcoming publication.

%\subsection{Application to the absorption system at $z=1.15$ towards HE0515$-$4414}

\section{Conclusions}
Historically, the list of fundamental constants has varied in time as our knowledge of physics improved and this trend is likely to continue. Searches for variations of ratios of dimensionless ratios of fundamental constants (with time, in space, matter density, or other parameters) are useful in this context because they lead to new insight. A new sample of high precision measurements will be made using new instrumentation (such as ESPRESSO) and new analysis methods (such as \texttt{AI-VPFIT}) to test the tentative detection of variations in the fine structure constant in quasar absorption systems. 

\section*{Acknowledgments}
DM thanks the organisers for the invitation to present at Moriond Gravitation Session. 

\section*{References}

%\begin{thebibliography}{99}
\bibliography{milakovic}
%\end{thebibliography}
%\printbibliography
\end{document}